\documentclass[12pt]{article}
\usepackage{newlfont}
\usepackage{amsmath}
\usepackage{times}
\usepackage[all]{xypic}
\usepackage{dcolumn}
\usepackage{bm}
\RequirePackage{graphicx}
\RequirePackage[figtopcap]{subfigure}
\usepackage[colorlinks]{hyperref}
\usepackage{authblk}
\usepackage{amssymb}
\usepackage{amsfonts}
\usepackage{yfonts}

\newcommand{\be}{\begin{equation}
\addtolength{\abovedisplayskip}{\extraspaces}
\addtolength{\belowdisplayskip}{\extraspaces}
\addtolength{\abovedisplayshortskip}{\extraspace}
\addtolength{\belowdisplayshortskip}{\extraspace}}
\newcommand{\ee}{\end{equation}}

\newcommand{\ba}{\begin{eqnarray}
\addtolength{\abovedisplayskip}{\extraspaces}
\addtolength{\belowdisplayskip}{\extraspaces}
\addtolength{\abovedisplayshortskip}{\extraspace}
\addtolength{\belowdisplayshortskip}{\extraspace}}
\newcommand{\ea}{\end{eqnarray}}

\numberwithin{equation}{section}
\begin{document}
\begin{center}
{\bf Isotropic stars in  higher-order  torsion scalar theories}
\end{center}
\begin{center}
{\bf Gamal G.L. Nashed}
\end{center}

\centerline{\it Center for Theoretical Physics, British University
in  Egypt} \centerline{\it Sherouk City 11837, P.O. Box 43, Egypt
\footnote{ Mathematics Department, Faculty of Science, Ain
Shams University, Cairo, 11566, Egypt \\
\hspace*{.2cm} Egyptian Relativity Group (ERG) URL:
http://www.erg.eg.net}}

\bigskip
\centerline{ e-mail:nashed@bue.edu.eg}
\hspace{2cm} \hspace{2cm}
\\
\\
\\
\\
\\
Two tetrad spaces reproducing spherically symmetric spacetime are applied to the  equations of motion of  higher-order torsion  theories.  Assuming the existence of conformal Killing vector, two isotropic solutions are derived. We show that the first solution is not stable while the second one confirms a stable behavior.  We also discuss the construction of the stellar model and show that one of our solution capable of such construction while the other cannot.  Finally, we discuss the generalized  Tolman-Oppenheimer-Volkoff and show that one of our models has a tendency to equilibrium.

\section{Introduction}

It is well known that in $f(T)$ gravity  inflation \cite{FF} and late time cosmic acceleration can be
realized in the early universe  \cite{Le0}--\cite{CDDS1}.  Recently there are many  models constructed to describe dark energy without the use of cosmological constant  (for more details  see review \cite{BCNO} and references therein). The  gravitational field equation  of $f(T)$ gravity is  second order as general relativity (GR).  $f(T)$ gravity suffers from non-invariance of local Lorentz transformation \cite{LSB}--\cite{KS}, non-minimal coupling
of teleparallel gravity to a scalar field \cite{GLSW}--\cite{GLSW2} and non-linear causality \cite{OINC}. Recently,  number of $f(T)$ gravitational theories have been proposed \cite{BF}--\cite{NBSP}. The structures of neutron and quark stars in $f(T)$ theory  have been discussed \cite{KHR}.  The anisotropic behavior, regularity conditions, stability and surface redshift of the compact stars have been checked \cite{AAZ}. Under those theories it is shown that $f(T)$ are not dynamically identical to teleparallel action plus a scalar field \cite{Yr11}. It has been shown that  investigations of $f(T)$, using observational data, are compatible with observations (see e.g. \cite{ZH,FF7} and references therein). A new
type of $f(T)$ theory  was proposed in order to explain the acceleration phase of the universe \cite{Yr}.  Also it has been shown that the well-known problem of frame dependence and violation of local Lorentz invariance in the formulation of $f(T)$ gravity is a consequence of neglecting the role of spin connection \cite{KS}.

$f(T )$ theory coupled with anisotropic fluid has been examined  for static spacetimes with spherical
symmetry and  many classes of solutions have been derived \cite{DRH2}. It has been shown that some conditions on the coordinates, energy density and
pressures, can produce new classes of anisotropic and isotropic solutions. Some of new black holes and wormholes solutions have been derived by selecting a set of non-diagonal tetrads \cite{DRH3}.  It has been   shown that relativistic stars can exist in $f(T)$  \cite{YLZ14}. A special analytic vacuum spherically symmetric solution with constant torsion scalar, within the framework of $f(T)$, has been derived \cite{Ngrg}. D-dimensional charged flat horizon solutions has been derived for a specific form of $f(T)$, i.e., $f(T)=T+\alpha T^2$ \cite{CGSV1}.  A complete investigation of the Noether symmetry approach in $f(T )$ gravity at FRW and spherical levels respectively has been investigated \cite{PBSCADT}. In the framework of $f(T)$ gravitational theories there are many solutions, spherically symmetric \cite{Nap}, spherically symmetric charged  \cite{Nprd}, homogenous anisotropic \cite{RHGR}, stability of the Einstein static closed and open universe \cite{LLG}. Some cosmological features of the $\Lambda$CDM model in the framework of the $f(T)$  are investigated \cite{SRKHT}.  However,  till now, no spherically symmetric isotropic   solution, {\it using non-diagonal tetrad fields},  derived in this theory.   It is the aim of the present study to find an analytic, isotropic spherically symmetric  solution   in higher-order torsion scalar theories. The arrangement of this study are as follows:  In Section \S 2, ingredients of $f(T)$ gravitational theory are provided. In Section \S 3, two different tetrad spaces having spherical symmetry are applied to the field equations of $f(T)$. Assuming the conformal Killing vector (CKV), we derived two non-vacuum spherically symmetric solutions in \S 3.  The physics relevant to the derived solutions are analyzed in \S 4. The energy conditions   are satisfied for the two solutions provided that the constants of integration be positive.  In addition, the stability condition, the nature of the star and Tolman-Oppenheimer-Volkoff (TOV) equation are   shown to be satisfied for one solution.  The results obtained in this study are discussed in final section.
\section{Ingredients of f(T) gravitational theory}
Another description of  Einstein's general relativity (GR) of gravitation is done through the employ  of what is called teleparallel equivalent of general relativity (TEGR). The ingredient quantity of this theory is the vierbein (tetrad) fields\footnote{ Greek letters $\alpha,\beta,...$ indicate spacetime indices while Latin indices $i,j,...$ run from $0$ to $3$ describe  Lorentz indices.} $\{h^{a}{_{\mu}}\}$ alternative to metric tensor fields $g_{\mu \nu}$. The associated metric  $g_{\mu\nu}=\eta_{ab}h^{a}{_\mu}h^{b}{_\nu}$ with $\eta_{ab}=\textmd{diag}(1,-1,-1,-1)$ being  Minkowskian metric, thus Levi-Civita symmetric connection $\mathring{\Gamma}^{\alpha}{_{\mu \nu}}$ is constructed from the metric and its first derivative \cite{MTW}. Within TEGR, it is possible to build a nonsymmetric connection,  Weitzenb\"{o}ck, $\Gamma^{\alpha}{_{\mu\nu}}=h^{a}{_{\mu}}\partial_{\nu}h_{a}{^{\alpha}}=-h_{a}{^{\alpha}}\partial_{\nu}h^{a}{_{\mu}}$.  Tetrad  space has a main merit that is  the null of the vierbein's   derivative, i.e. $\nabla_{_{\nu}}h^{a}{_{\mu}}\equiv 0$, where  $\nabla$, regarding the nonsymmetric  Weitzenb\"{o}ck connection. Therefore, the vanishing of the vierbein's covariant derivative recognizes  auto-parallelism or absolute parallelism condition. Actually, the $\nabla$ operator is not invariant under local Lorentz transformations (LLT).  The metric $g_{\mu \nu}$  is not able to guess  one set of vierbein fields; thus   extra  freedom  need to be determined so as to determine unique  frame. Because of the absolute parallelism condition, it can be shown that the metricity condition is satisfied. The Weitzenb\"{o}ck connection is curvatureless while it has a non vanishing torsion tensor $T$ given as
\begin{equation}\label{torsion}
{T^\lambda}_{ \mu \nu} := {h_a}^{\lambda}(\partial_\mu {h^a}_{\nu}-\partial_\nu  {h^a}_{\mu}),
\end{equation}
and contortion tensor $K$
\begin{equation}\label{contortion}
{K^{\mu \nu}}_\alpha =
-\frac{1}{2}\left({T^{\mu \nu}}_\alpha-{T^{\nu
\mu}}_\alpha-{T_\alpha}^{\mu \nu}\right).
\end{equation}
 The torsion scalar of TEGR theory is given by
\begin{equation}\label{Tor_sc}
 \mathrm{T} := {T^\alpha}_{\mu \nu} {S_\alpha}^{\mu \nu},
\end{equation}
with  $S$ defined as
\begin{equation}\label{Stensor}
{\mathrm{S}_\alpha}^{\mu \nu} := \frac{1}{2}\left({\mathrm{K}^{\mu\nu}}_\alpha+\mathrm{\delta}^\mu_\alpha{\mathrm{T}^{\beta
\nu}}_\beta-\mathrm{\delta}^\nu_\alpha{\mathrm{T}^{\beta \mu}}_\beta\right).
\end{equation}
Equation (\ref{Stensor}) shows skewness   in $\mu$ and $\nu$. Like $f(R)$, we could establish  Lagrangian of $f(T )$ like 
\begin{eqnarray}\label{q7}
& & {\cal \mathrm{L}}({\mathrm{h}^a}_\mu, \mathrm{\Phi}_A)=\int
d^4x h\left[\frac{1}{16\pi}f(\mathrm{T})+{\cal L}_{Matter}(\mathrm{\Phi}_A)\right], \quad \textrm
{where} \quad \mathrm{h}=\mathrm{\sqrt{-g}}=det\left({\mathrm{h}^a}_\mu\right), \nonumber\\
& & \mathrm{\Phi}_A \quad \textrm{are}  \quad \textrm{ matter} \quad  \textrm {fields}.
\end{eqnarray}
In this study we  postulate  the units in which $G = c = 1$.   Lagrangian (\ref{q7}) can consider as a function of the fields ${h^a}_\mu$. Variation of Lagrangian  (\ref{q7}) with respect to  the tetrad field ${h^a}_\mu$  we obtain the following field equations  \cite{BF,CGSV1} 
\begin{equation}\label{q8}
{S_\mu}^{\rho \nu}\, T_{,\rho} \
f(T)_{TT}+\left[h^{-1}{h^a}_\mu\partial_\rho\left(h{h_a}^\alpha
{S_\alpha}^{\rho \nu}\right)-{T^\alpha}_{\lambda \mu}{S_\alpha}^{\nu
\lambda}\right]f(T)_T-\frac{1}{4}\delta^\nu_\mu f(T)=-4\pi {\mathfrak{T}_{\mu}}^{\nu},
\end{equation}
where
$T_{,\rho}=\displaystyle\frac{\partial T}{\partial x^\rho}$, \; \; $\mathrm{f(T)_T}=\displaystyle\frac{\partial \mathrm{f(T)}}{\partial T}$, \; \;$\mathrm{f(T)_{TT}}=\displaystyle\frac{\partial^2 \mathrm{f(T)}}{\partial T^2}$ and  ${\mathfrak{T}_{\mu}}^{\nu}$ denotes the energy-momentum tensor of the anisotropic
fluid which is defined as
\begin{equation}\label{em}
{\mathfrak{T}_{\mu}}^{\nu}=(\rho+p_t)u_\mu u^\nu-p_t{\delta_\mu}^\nu+(p_r-p_t)\eta_\mu\eta^\nu,\end{equation}
with $p_r$ represents  the radial pressure, $p_t$   the tangential pressure and
\begin{equation}u_\mu u^\mu=-\eta_\mu \eta^\mu=1, \qquad {and} \qquad u^\mu \eta_\mu=0.\end{equation}
 Equations (\ref{q8}) are the field equations of$f(T)$ gravitational theory.

\section{Non-vacuum spherically symmetric solutions in higher-order torsion scalar theories}
In this section, we are going to apply two, non-diagonal, different tetrad fields having spherical symmetry to the field equations (\ref{q8}).

\subsection{First tetrad}

The  equation of motion of GR supply   rich
field to use symmetries which link geometry
and matter in a natural way. Collineations are  symmetries which come
either from geometrical viewpoint  or physical relevant
quantities. The importance  of  collineations is the CKV which provides a more information of the
construction of the spacetime geometry. The employs of the CKV   simplifies  the equations of motions of GR. The CKV is defined as
\begin{equation} \label{kv}
{\mathbf L}_\zeta g_{i j}=\zeta_{i;j}+\zeta_{j;i}=\psi g_{i j},
\end{equation}
with  ${\mathbf L}$ being the  Lie derivative  
and the $\psi$ being the conformal factor. One can assume
the vector $\zeta$ which creates the conformal symmetry and
makes the metric  conformally mapped onto itself through
$\zeta$. One must  note that  $\zeta$ and $\psi$   not necessary
 be static even supposing a static metric
\cite{BHL}. In addition, one must
be careful about the following: \vspace{0.2cm}\\ (i) if $\psi = 0$, then (\ref{kv}) leads to  a Killing
vector,\vspace{0.2cm}\\ (ii) if $\psi=constant$, then (\ref{kv}) leads to  homothetic vector \vspace{0.2cm}\\ (iii) if $\psi =\mathrm{ \psi(x, t)}$ then (\ref{kv}) leads to conformal vectors.
Furthermore, if  $\psi$  is vanishing then the spacetime behaves as
asymptotically flat and one has a null Weyl
tensor. Thus, to have more understanding of the
spacetime geometry one must take into account the CKV. Essentially, the Lie derivative  
${\mathbf L}$ shows the inner  field of gravity of a
stellar configuration related to the vector field $\zeta$.

The first tetrad  field having a   spherical
symmetry takes the shape \cite{Nmpla}
\begin{equation} \label{te1}
\begin{tabular}{l}
\hspace{-0.3cm}
$\left( {h^i}_\mu \right)=$ \\[3pt]
  $\left(
  \begin{array}{cccc}
\displaystyle\frac{{\cal F}_1(r)}{{\cal F}_2(r)} &{\cal F}_2(r)& 0 &
0\\[9pt] \sin\theta
\cos\phi &{\cal F}_1(r)\sin\theta
\cos\phi&r\cos\theta \cos\phi & -r\sin\theta \sin\phi \\[9pt]  \sin\theta
\sin\phi&{\cal F}_1(r)\sin\theta \sin\phi&r\cos\theta
\sin\phi & r\sin\theta \cos\phi \\[9pt]  \cos\theta
 & {\cal F}_1(r)\cos\theta&-r\sin\theta & \\[9pt]
 \end{array}
\right)$,
\end{tabular}
\end{equation}
 where ${\cal F}_1(r)$,  and  ${\cal F}_2(r)$ are two unknown
 functions of the radial coordinate, $r$.

The associated metric of (\ref{te1}) takes the form
\begin{equation} \label{m1}
ds^2=-\frac{{{\cal F}_1}^2-{{\cal F}_2}^2}{{{\cal F}_2}^2}dt^2+({{\cal F}_1}^2-{{\cal F}_2}^2)dr^2+d\Omega, \qquad d\Omega=r^2(d\theta^2+\sin^2\theta d\phi^2),
\end{equation}
which is a static spherically
symmetric spacetime admits  one parameter group
of conformal motion.
Equation (\ref{m1})
is conformally mapped onto itself along $\zeta$.
Therefore, (\ref{kv}) leads to
\begin{eqnarray} \label{m2} && 2[{\cal F'}_1{{\cal F}_1}{{\cal F}_2}-{{\cal F}_1}^2{{\cal F'}_2}]\zeta^1=\psi(r)[ {{\cal F}_1}^2{\cal F}_2-{{\cal F}_2}^3], \nonumber\\
&& \zeta^0=c, \qquad \qquad \qquad \zeta^1=\frac{\psi(r) r}{2},\nonumber\\
&&   2\zeta^1[{\cal F}_1 {\cal F'}_1-{\cal F}_2 {\cal F'}_2]+2\zeta'^1[{{\cal F}_1}^2-{{\cal F}_2}^2]=\psi(r)[{{\cal F}_1}^2-{{\cal F}_2}^2], \end{eqnarray}
where 0 and 1 refer to the temporal  and  spatial coordinates. Equation  (\ref{m2}) leads to
\begin{equation}  \label{so1}  {\cal F}_1=\frac{\sqrt{1+{c_0}^2r^2}{\cal F}_3}{c_0 r}, \qquad  {\cal F}_2=\frac{{\cal F}_3}{rc_0}, \qquad \zeta^i=c_1{\delta^i}_0+\frac{\psi(r) r}{2}{\delta^i}_1, \qquad {{\cal F}_1}\neq{{\cal F}_2}, \qquad {{\cal F}_3}=\frac{c_2}{\psi(r)},\end{equation}
with $c$, $c_0$, $c_1$ and $c_2$ are  constants of integration.

Using (\ref{so1}), tetrad  (\ref{te1}) is rewritten as
\begin{equation} \label{te2} \hspace{-0.3cm}\begin{tabular}{l}
  $\left( {h^i}_\mu \right)=$\\[3pt]
  $\left(
  \begin{array}{cccc}

   \sqrt{1+{c_0}^2r^2} & \displaystyle\frac{{\cal F}_3}{rc_0}& 0 &
0\\[9pt] \sin\theta
\cos\phi &\displaystyle\frac{{\cal F}_3\sqrt{1+{c_0}^2r^2}\sin\theta
\cos\phi}{rc_0}&r\cos\theta \cos\phi & -r\sin\theta \sin\phi \\[9pt]
 \sin\theta
\sin\phi&\displaystyle\frac{{\cal F}_3\sqrt{1+{c_0}^2r^2}\sin\theta \sin\phi}{rc_0}&r\cos\theta
\sin\phi & r\sin\theta \cos\phi \\[9pt] \cos\theta  &
\displaystyle\frac{{\cal F}_3\sqrt{1+{c_0}^2r^2}\cos\theta}{rc_0}&-r\sin\theta & 0 \\[9pt]
  \end{array}
\right)$,
\end{tabular}\end{equation}

Tetrad field (\ref{te2}) has the following associated metric \begin{equation}
ds^2=-{c_0}^2r^2dt^2+{{\cal F}_3}^2dr^2+d\Omega, \qquad d\Omega=r^2(d\theta^2+\sin^2\theta d\phi^2),
\end{equation}
Using (\ref{te2}) in (\ref{Tor_sc}) we get  
\begin{equation} \label{ts}
\mathrm{T}=2\frac{2(1+2{c_0}^2r^2){\cal F}_3(r)-rc_0\sqrt{1+{c_0}^2r^2}(3+{{\cal F}_3}^2(r))}{c_0r^3\sqrt{1+{c_0}^2r^2}{{\cal F}_3}^2(r)},\end{equation}
Using Eqs. (\ref{ts}) and (\ref{te2}) in the field equations (\ref{q8}) we get the following non-vanishing components:
\begin{eqnarray}  \label {fe} & & 4\pi {{\cal T}_0}^0=4\pi \rho=-\frac{\sqrt{1+{c_0}^2r^2}{\cal F}_3-rc_0}{r^2c_0{{\cal F}_3}^2}T'f_{TT}\nonumber\\
& &+\frac{rc_0\sqrt{1+{c_0}^2r^2}(2{{\cal F}_3}-r{\cal F'}_3)-{{\cal F}_3}^2(1+2{c_0}^2r^2)}{r^3c_0\sqrt{1+{c_0}^2r^2}{{\cal F}_1}^3}f_{T}+\frac{f}{4},\nonumber\\
& & 4\pi {{\cal T}_1}^0=-\frac{T'f_{TT}}{r^3{c_0}^2},\nonumber\\
& & -4\pi {{\cal T}_1}^1=4\pi p_r=-\frac{{{\cal F}_3}(1+2r^2{c_0}^2)-3rc_0\sqrt{1+{c_0}^2r^2}}{r^3c_0{{\cal F}_1}^2\sqrt{1+{c_0}^2r^2}}f_{T}+\frac{f}{4},\nonumber\\
& & -4\pi {{\cal T}_2}^2=-4\pi {{\cal T}_3}^3=4\pi p_t=-\frac{\sqrt{1+{c_0}^2r^2}{\cal F}_3-2rc_0}{2r^2c_0{{\cal F}_3}^2}T'f_{TT}\nonumber\\
& & +\frac{rc_0\sqrt{1+{c_0}^2r^2}({{\cal F}_3}^3+4{{\cal F}_3}-2r{\cal F'}_3)-2{{\cal F}_3}^2(1+2{c_0}^2r^2)}{2r^3c_0\sqrt{1+{c_0}^2r^2}{{\cal F}_3}^3}f_{T}+\frac{f}{4}.
\end{eqnarray}
Second equation of (\ref{fe}) leads to  $f_{TT}= 0$, or $T=constant$. The case $T=constant$ gives a constant function and this is out the scope of the present study.  Therefore, we  seeking solutions make constrain on the form of $f(T)$ to have the form
\begin{equation} \label{ft} f(T)=T, \qquad \quad \Rightarrow f_{TT}=0.\end{equation}
 Assuming the isotropic condition
\begin{equation}\label{co} p_r=p_t=p,\end{equation}
and using (\ref{co}) in  (\ref{fe}), we get:
\begin{eqnarray} \label{co1}
& & {\cal F}_3(r)=\frac{2}{\sqrt{2+4r^2c_3}},\nonumber\\
& & T=-\frac{6c_3c_0r^3\sqrt{1+r^2{c_0}^2}+5rc_0\sqrt{1+r^2{c_0}^2}-2(1+2{c_0}^2r^2)\sqrt{2+4r^2{c_3}}}
{r^3c_0\sqrt{1+r^2{c_0}^2}},\nonumber\\
& & 16 \pi \rho=\frac{6c_3r^2-1}{r^2}, \qquad  16 \pi p=\frac{6c_3r^2+1}{r^2}, \qquad  \psi(r)=\frac{c_2\sqrt{2+4r^2c_3}}{2}.
\end{eqnarray}
The sound velocity ${v_s}^2$  is defined as ${v_s}^2:=\frac{dp}{d\rho}$. Using (\ref{co1}) we get the sound velocity in the form
\begin{equation}
{v_s}^2=-1.\end{equation}
\subsection{Second tetrad}
The second tetrad space having a stationary and spherical
symmetry takes the form \cite{Nprd} \begin{equation}
 \label{te3}
\hspace{-0.3cm}\begin{tabular}{l}
  $\left( {h^i}_\mu \right)=$\\[5pt]
  $\left(
  \begin{array}{cccc}
  {\cal F}_4(r) &0& 0 &
0\\[9pt] 0 &{\cal F}_5(r)\sin\theta
\cos\phi&r\cos\theta \cos\phi & -r\sin\theta \sin\phi \\[9pt] 0&{\cal F}_5(r)\sin\theta \sin\phi&r\cos\theta
\sin\phi & r\sin\theta \cos\phi \\[9pt]0  &
{\cal F}_5(r)\cos\theta&-r\sin\theta & 0   \\[9pt]
  \end{array}
\right)$,
\end{tabular}\end{equation}
 where ${\cal F}_4(r)$   and  ${\cal F}_5(r)$ are unknown
 functions. Using the same procedure applied to tetrad (\ref{te1}) we get the following equations of CKV of tetrad (\ref{te3})
\begin{equation} 2{\cal F}'_4\xi^1=\psi(r) {\cal F}_4, \qquad \xi^0=c, \qquad \xi^1=\frac{\psi(r) r}{2}, \qquad 2\xi^1{\cal F}'_5+2{\xi^1}_{,1}{\cal F}_5=\psi(r){\cal F}_5, \end{equation}
 The above set of equations imply
\begin{equation} \label{so5} {\cal F}_4=c_4 r, \qquad  {\cal F}_5=\frac{c_5}{\psi(r)}, \qquad \xi^i=c_6{\delta^i}_0+\frac{\psi(r) r}{2}{\delta^i}_1.\end{equation}
where $c_4$, $c_5$ and $c_6$  are constants of integration .

Using  (\ref{so5}), tetrad  (\ref{te3}) can be rewritten as
\begin{equation} \label{te4}
\hspace{-0.3cm}\begin{tabular}{l}
  $\left( {h^i}_\mu \right)=$\\[5pt]
  $\left(
  \begin{array}{cccc} c_4r & 0& 0 &
0\\[9pt] 0 &{\cal F}_5\sin\theta
\cos\phi&r\cos\theta \cos\phi & -r\sin\theta \sin\phi \\[9pt] 0&{\cal F}_5\sin\theta \sin\phi&r\cos\theta
\sin\phi & r\sin\theta \cos\phi \\[9pt] 0  &
{\cal F}_5\cos\theta&-r\sin\theta & 0 \\[9pt]
  \end{array}
\right)$.
\end{tabular}\end{equation}
Using (\ref{te4}), the torsion scalar  (\ref{Tor_sc}), takes the form
\begin{equation}\label{ts3} T=\frac{2(3-4{\cal F}_5+{{\cal F}_5}^2)}{r^2{{\cal F}_5}^2}.\end{equation}
Inserting (\ref{ts3}) and the components of the tensors ${S^{\nu \mu}}_\rho$ and  ${T^{\nu \mu}}_\rho$ in the field equations (\ref{q8}) we obtain
\begin{eqnarray} \label{fe3} & & 4\pi {{\cal T}_0}^0=4\pi\rho=\frac{(1-{{\cal F}_5})}{r{{\cal F}_5}^2}T'f_{TT}-\frac{2{{\cal F}_5}^2-2{\cal F}_5+r{\cal F}'_5}{r^2{{\cal F}_5}^3}f_{T}+\frac{f}{4},\nonumber\\
& &- 4\pi {{\cal T}_1}^1=4\pi p_r=\frac{3-2{\cal F}_5}{r^2{{\cal F}_5}^2}f_{T}+\frac{f}{4},\nonumber\\
& & -8\pi {{\cal T}_2}^2= -8\pi {{\cal T}_3}^3=4\pi p_t=\frac{(2-{{\cal F}_5})}{2r{{\cal F}_5}^2}T'f_{TT}-\frac{2{{\cal F}_5}^2-2{\cal F}_5+r{\cal F}'_5-{{\cal F}_5}^3}{2r^2{{\cal F}_5}^3}f_{T}+\frac{f}{4},\nonumber\\
& &
\end{eqnarray}
The above system cannot be solved without assuming some a specific constraint on the form of $f(T)$.  Therefore, we are going to use the constraint (\ref{ft}) in  (\ref{fe3}) and obtain the following
\begin{eqnarray} \label{so3}
& & {\cal F}_5(r)=\frac{2}{\sqrt{2+4r^2c_7}},\nonumber\\
& & T=\frac{5\sqrt{2+4r^2c_7}-8-16r^2c_5+6c_7\sqrt{2+4r^2c_7}}{r^2\sqrt{2+4r^2c_7}},\nonumber\\
& & 16 \pi \rho=\frac{9+18r^2c_7-8\sqrt{2+4r^2c_7}}{r^2}, \qquad 16 \pi p=\frac{11+18r^2c_7-8\sqrt{2+4r^2c_7}}{r^2},\nonumber\\
& & \psi(r)=\frac{c_5\sqrt{2+4r^2c_7}}{2}.
\ea
Using (\ref{so3}), the sound velocity ${v_s}^2$ takes the form
\begin{equation}
\frac{dp}{d\rho}=\frac{16+32r^4{c_7}^2+48r^2c_7-22r^2c_7\sqrt{2+4r^2c_7}-11\sqrt{2+4r^2c_7}}
{(8c_7\sqrt{2+4r^2c_7}-18r^2c_7-9+8\sqrt{2+4r^2c_7})\sqrt{2+4r^2c_7}}.\end{equation}

\section{Physics relevant to the models}
\underline{\bf Energy conditions:}\vspace{0.2cm}\\
 Energy conditions are essential tools   to understand  cosmology   
and general results related to   strong gravitational
fields. These tools are  three energy conditions,
null energy  (NEC), the strong energy (SEC)  and weak
energy conditions (WEC) \cite{HE}--\cite{ZW}. Such conditions  have the  following inequalities
\begin{eqnarray}\label{fe4}
& & NEC:   \rho+p_r\geq 0, \qquad \rho+p_t\geq 0,\nonumber\\
& &  SEC: \rho+p_r\geq 0, \qquad \qquad \rho+p_r+2p_t\geq 0,\nonumber\\
& & WEC: \rho\geq 0, \qquad \qquad \rho+p_r\geq 0,\qquad \qquad \rho+p_t\geq 0.\end{eqnarray}

The  broken of (\ref{fe4}) leads to   ghost instabilities.\vspace{0.2cm}\\
\underline{\bf Energy conditions of smooth transition  models }\vspace{0.2cm}\\
Let us apply the above procedure of the energy conditions  given by (\ref{fe4}) to the derived solutions given in the previous section.
For the case of isotropic, i.e., $p_r=p_t$, we can see from  figures 1 and 2:
The density  has a positive value and the conditions  $\rho+p\geq 0 $ $\rho+3p\geq 0 $ are satisfied when the constant  $c_3>0$ for the first model  and $c_7>0$ for the second model. This  means that NEC, SEC and WEC conditions are satisfied for the above two models. Also it is interesting to note that the density and pressure of both solutions do not depend on the constants $c_0$ and $c_4$. \vspace{0.2cm}\\
\begin{figure}
\centering
\subfigure[figtopcap][$\rho$ ]{\label{fig1a}\includegraphics[scale=.25]{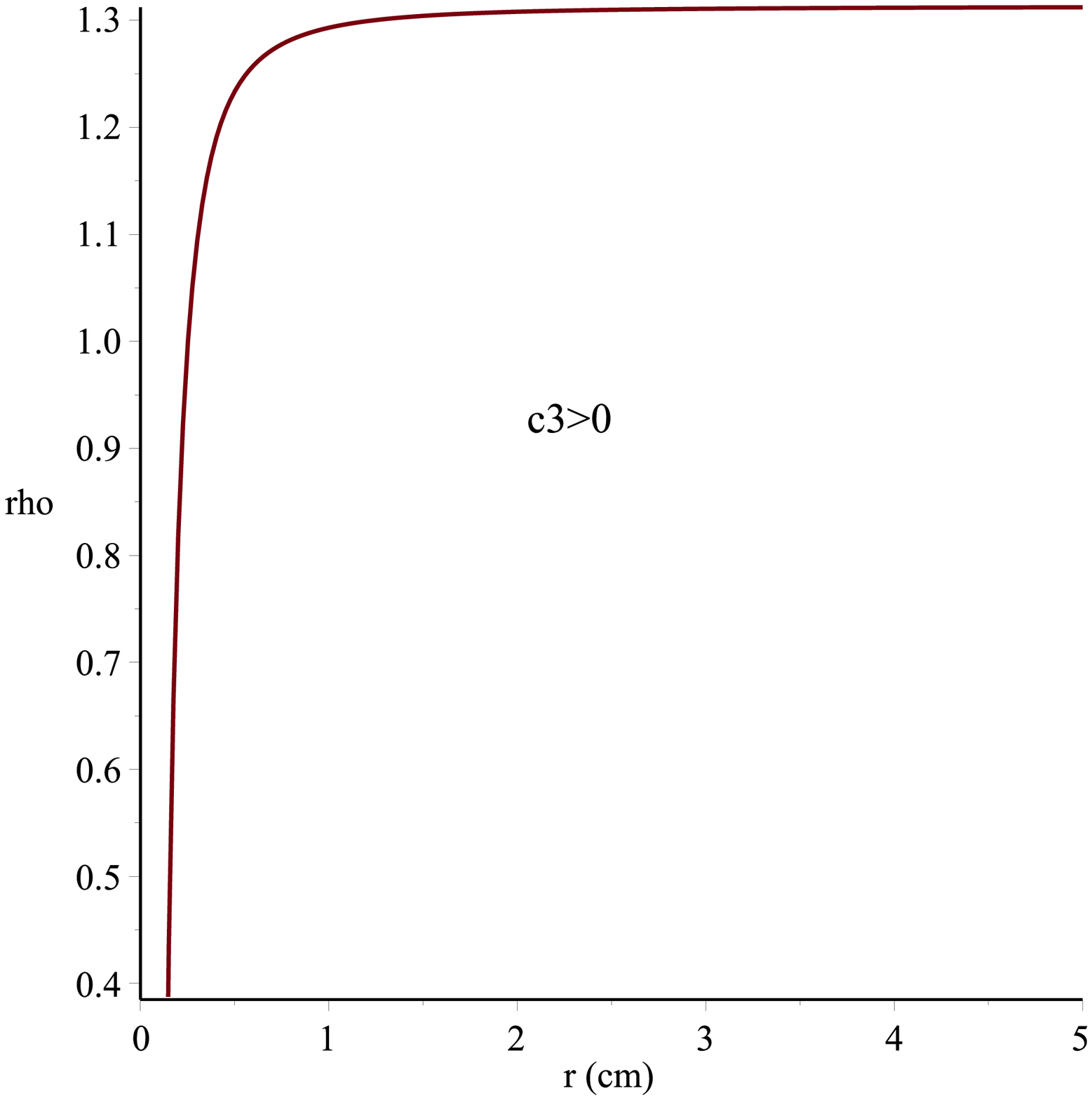}}
\subfigure[figtopcap][$\rho$+p   ]{\label{fig1b}\includegraphics[scale=.25]{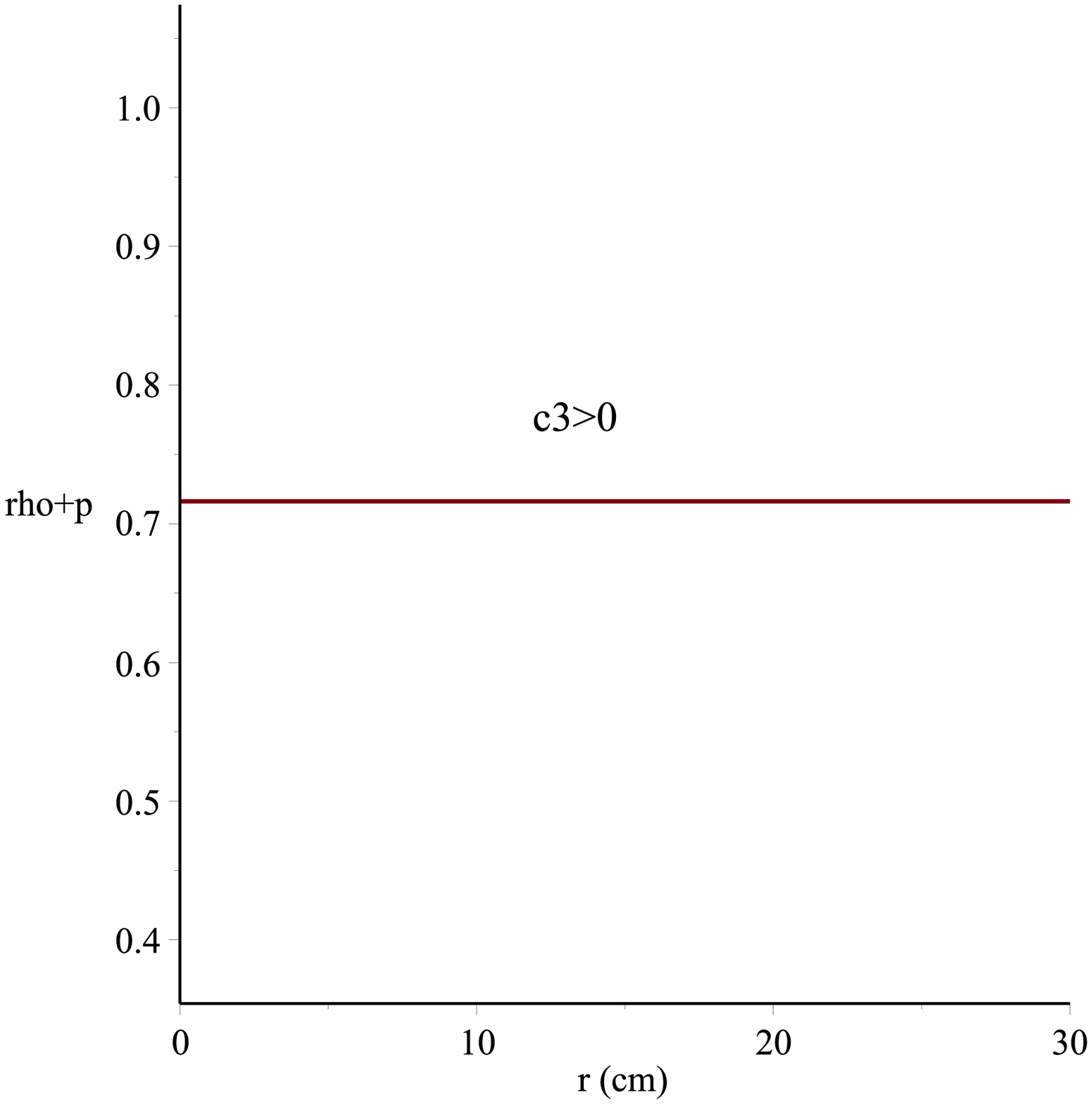}}
\subfigure[figtopcap][$\rho$+3p   ]{\label{fig1c}\includegraphics[scale=.25]{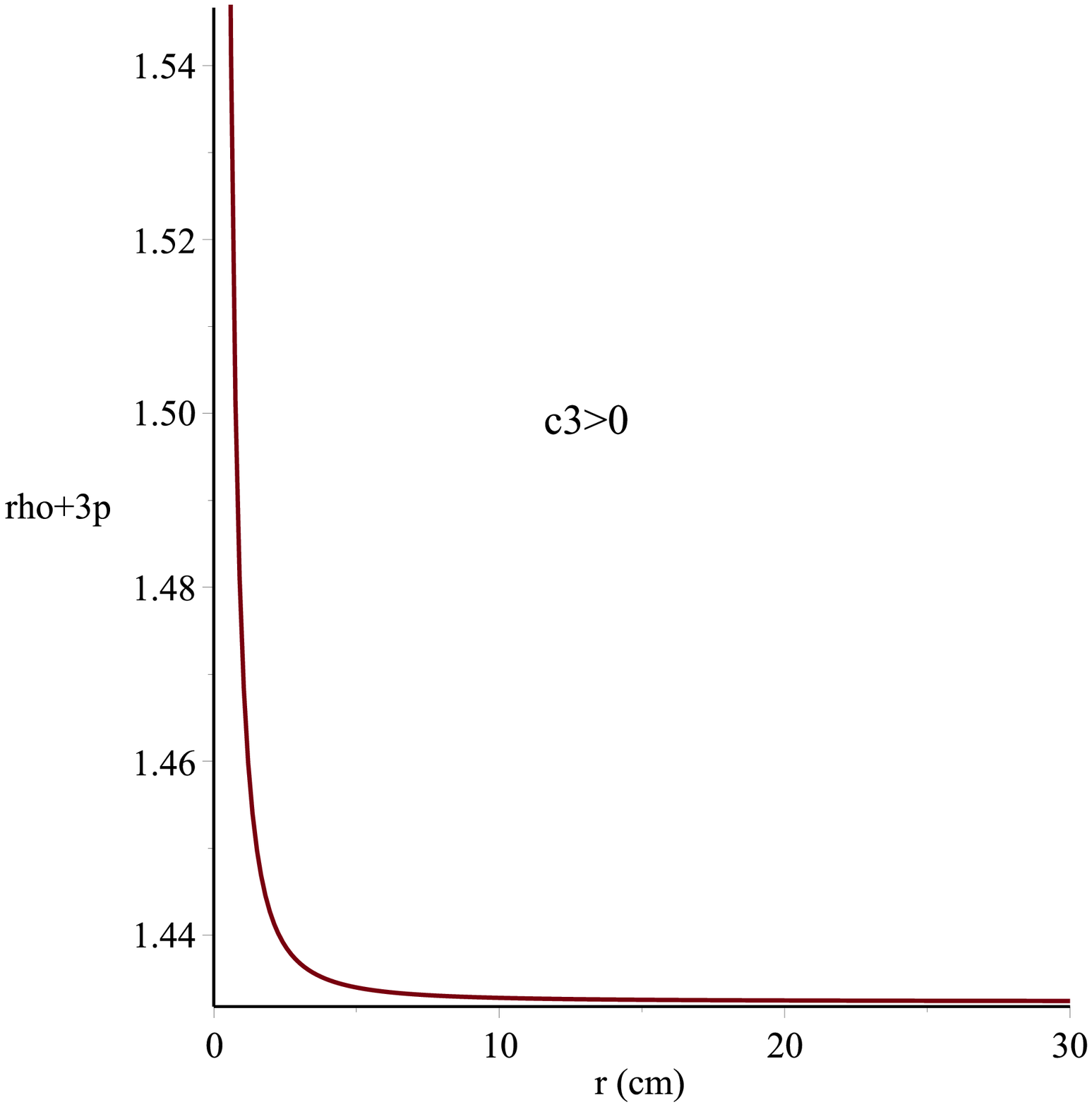}}
\caption{Energy conditions of the first model: The constant $c_3$ assumes a positive value so that the density be positive and also the pressure.}
\label{Fig1}
\end{figure}
\begin{figure}
\centering
\subfigure[figtopcap][$\rho$ ]{\label{fig2a}\includegraphics[scale=.25]{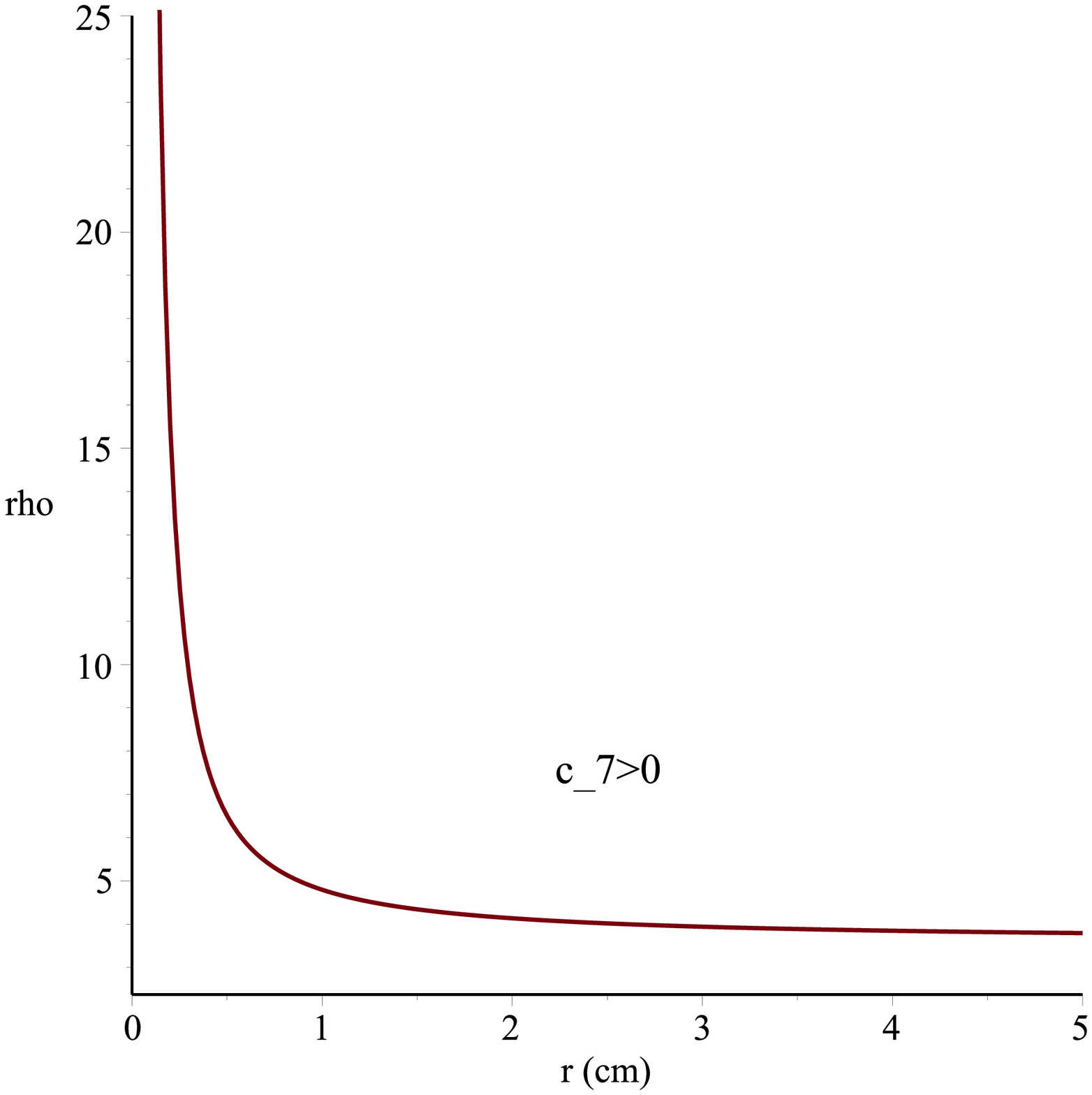}}
\subfigure[figtopcap][$\rho$+p   ]{\label{fig2b}\includegraphics[scale=.25]{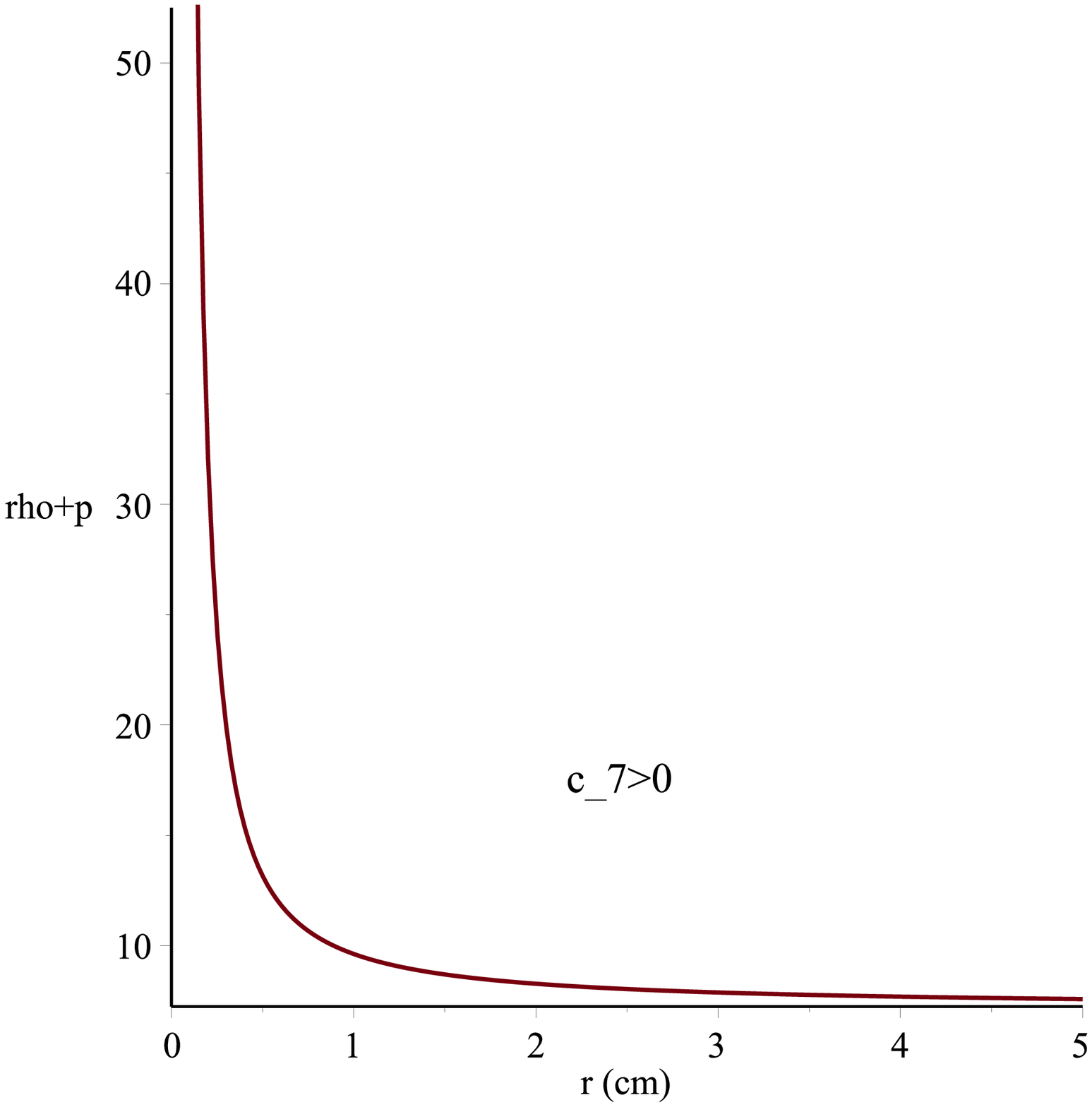}}
\subfigure[figtopcap][$\rho$+3p   ]{\label{fig2c}\includegraphics[scale=.25]{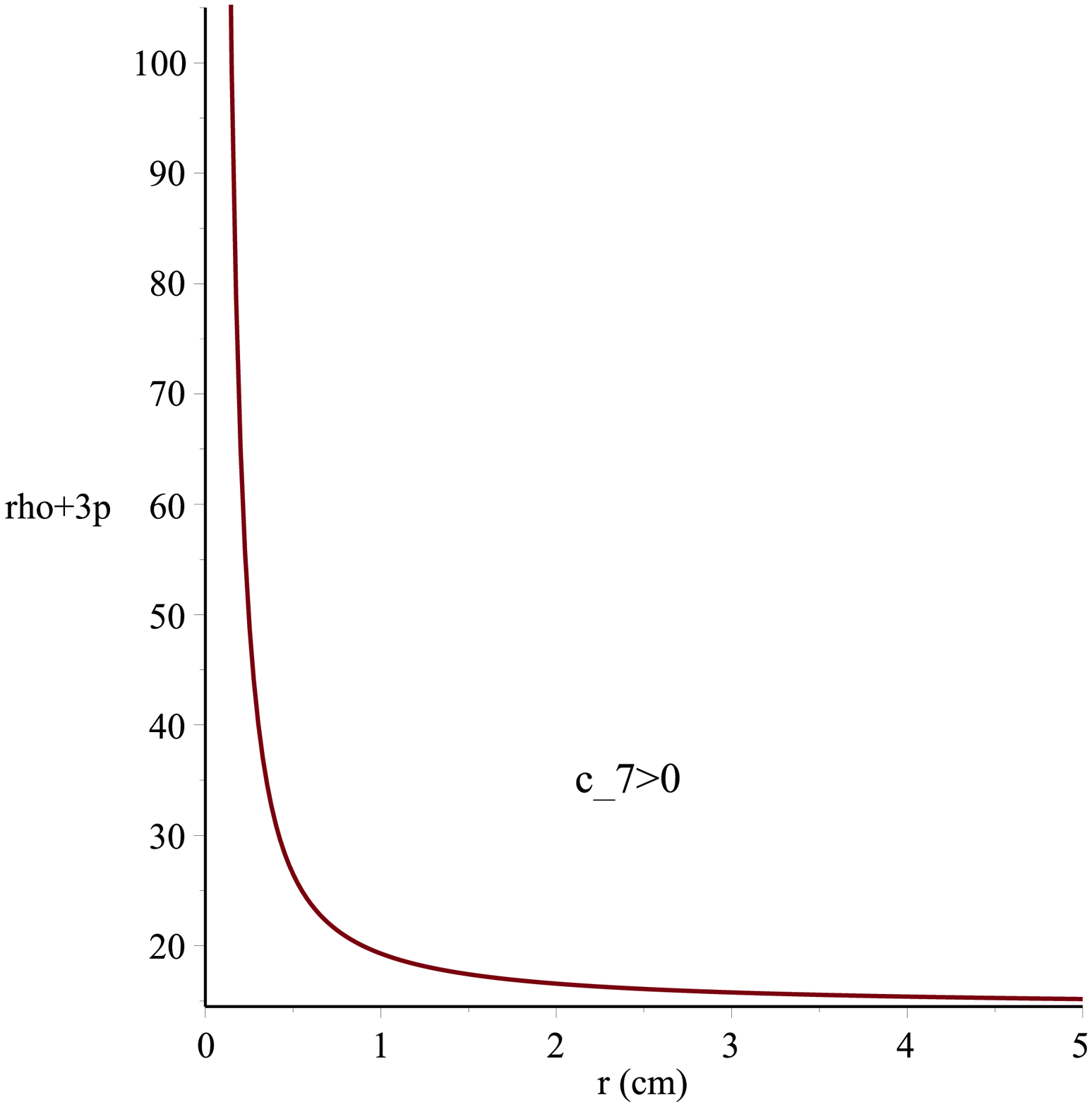}}
\caption{Energy conditions of the second model: The constant $c_7$ assumes a positive value so that the density be positive and also the pressure.}
\label{Fig1}
\end{figure}
\underline{\bf Stability problem}\vspace{0.2cm}\\
To study the stability issue
of the above two models we use the cracking mechanism \cite{Hl} in which the squared of speed  sound 
must lies in the range [0,\, 1], i.e., $0\leq {v_s}^2 \leq 1$.  Figure 3 (a) does not show the positivity
criterion i.e., ${v_s}^2\leq0$. However, Fig. 3 (b) satisfies the criterion of stability i.e., ${v_s}^2 \geq0$
within the matter distribution provided any value of the constant $c_7$, in figure 3 (b),
and thus second solution preserves stability.\vspace{0.2cm}\\
\begin{figure}
\centering
\subfigure[figtopcap][${v_s}^2<0$ ]{\label{fig3a}\includegraphics[scale=.25]{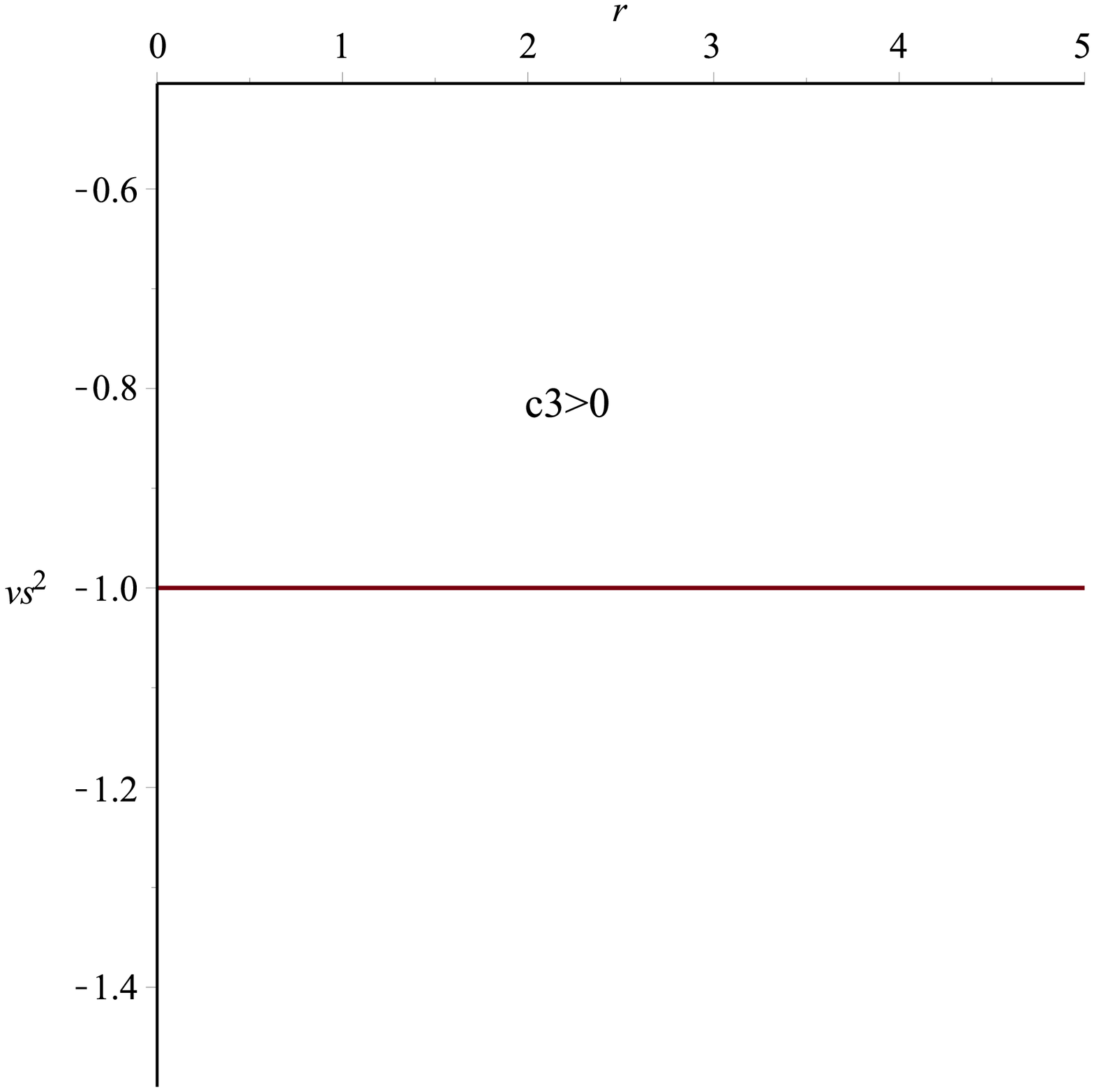}}
\subfigure[figtopcap][${v_s}^2\geq0$   ]{\label{fig3b}\includegraphics[scale=.25]{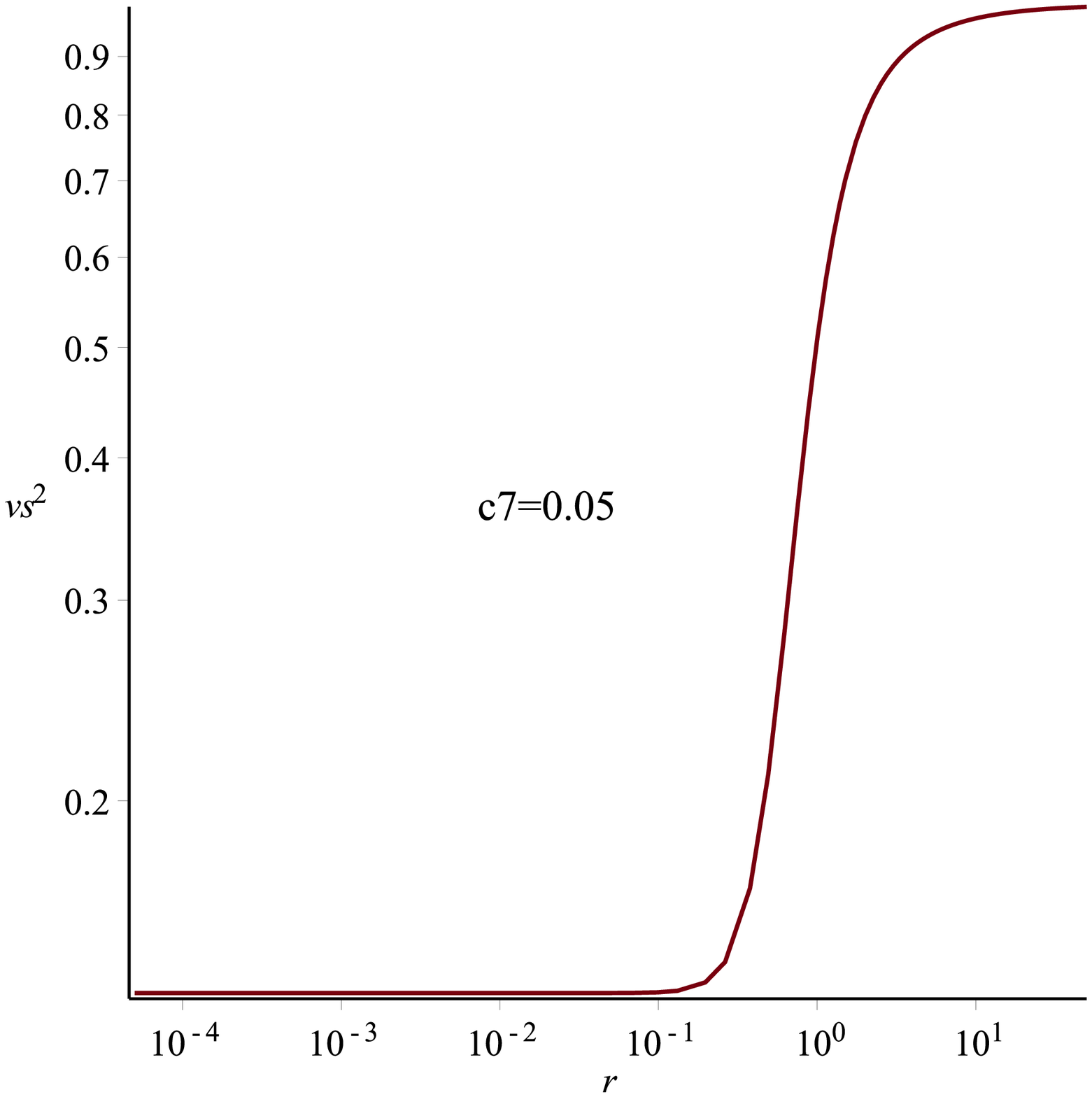}}
\caption{(a) shows that the first model does not has stability because the sound velocity ${v_s}^2 \notin [0,1]$ as is required while  (b) shows a stability  behavior.}
\label{Fig3}
\end{figure}
\underline{\bf Nature of the star}\vspace{0.2cm}\\
To understand the star behavior we use a plot to indicate the radius of the stellar
model for the second solution. Fig. 4, shows the cut on r-axis
is approximately 1 km (Fig. 4). 
This value is a  small value and shows  a compact
star \cite{BRRC,DRGR}.  
The value  $R\sim 1$ km  produces
us to find the surface density of the system.
As $r$ approximately vanishing, density approximates   $\infty$ and thus,
 the central density is far from the aim of the present  study. Only, we can inspect the surface density
by close the values of the Newtonian constant, $G$, and the speed of light, $c$ in the expression
of density  which gives  the numerical value as
$13 gm/cm^3$. This is a normal energy density
in which the  radius $R =
1 km$ is very small.  This shows that the second solution of $f(T)$
describes an ultra-compact star \cite{Rr}--\cite{HR}. The first model is not a physical one because to find the cutting of the pressure with the r-axis the constant $c_3<0$ which produces a contradiction with the energy conditions.
\begin{figure}
\centering
\subfigure[figtopcap][ p cut of the second model  ]{\label{fig4}\includegraphics[scale=.25]{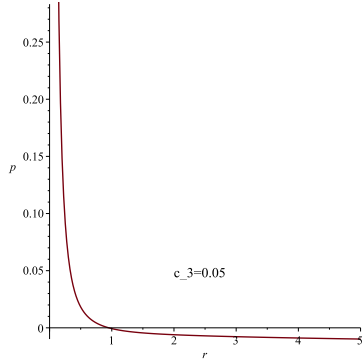}}
\caption{The radius of  star is shown when the pressure  cuts r-axis.}
\label{Fig4}
\end{figure}\vspace{0.1cm}\\
\underline{\bf TOV equation}\vspace{0.2cm}\\
The TOV
equation has the shape \[ ds^2=-e^{\nu(r)}dt^2+e^{\lambda(r)}dr^2+r^2(d\theta^2+sin^2\theta d\phi^2), \] can be written in the form \cite{BRRC}
\begin{equation} \label{tov}
-\frac{M_{G(r)}(\rho+p_r)e^{\frac{\lambda(r)-\nu(r)}{2}}}{r^2}-\frac{dp_r}{dr}+\frac{2(p_t-p_r)}{r}=0,
\end{equation}
where $M_{G(r)}$ is mass of gravity in  a 
sphere with radius $r$ which has the form
\begin{equation} \label{gm}
M_{G(r)}=\frac{r^2 \nu'e^{\frac{\lambda(r)-\nu(r)}{2}}}{2}.\end{equation}
Using (\ref{gm}) in   (\ref{tov}), we obtain in the isotropic case
\begin{equation} \label{tov1}
-\frac{\nu'(\rho+p_r)}{2}-\frac{dp_r}{dr}=0,
\end{equation}
Equation (\ref{tov1}) demonstrates the equilibrium
of the  configuration under 
 distinct  forces. As an equilibrium condition we  write (\ref{tov1})
in the   form:
\begin{equation}
F_g+F_h=0,
\end{equation}
where
\begin{equation}
F_g=-\frac{\nu'(\rho+p_r)}{2}, \qquad \qquad F_h=-\frac{dp_r}{dr}. \end{equation}
Using (\ref{te2}),  (\ref{co1}),  (\ref{te4}) and (\ref{so3})  we  plot the feature of TOV equation for
the above two models in Figure 5.
\begin{figure}
\centering
\subfigure[figtopcap][TOV of the first model ]{\label{fig5a}\includegraphics[scale=.25]{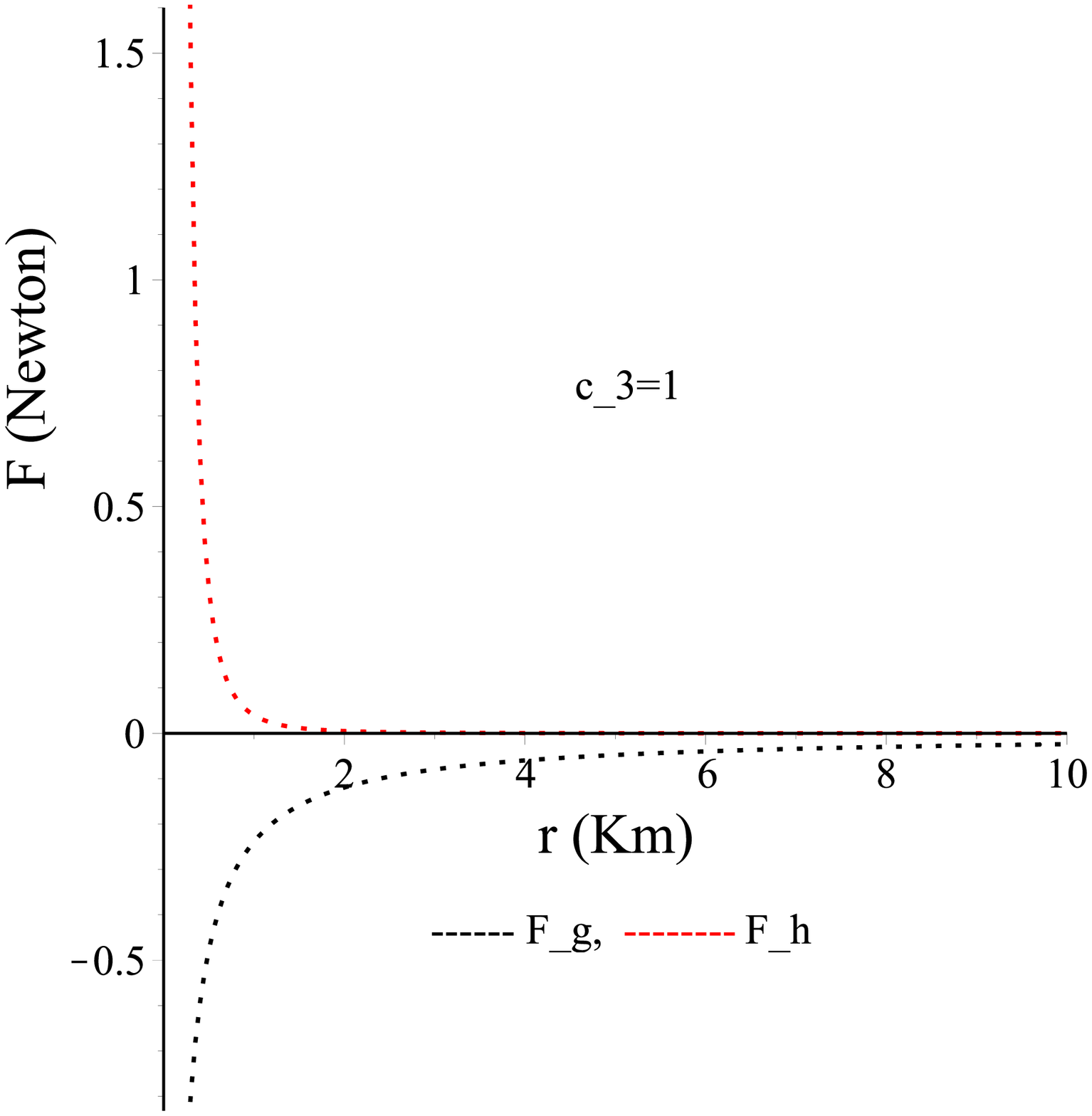}}
\subfigure[figtopcap][TOV of the second model  ]{\label{fig5b}\includegraphics[scale=.25]{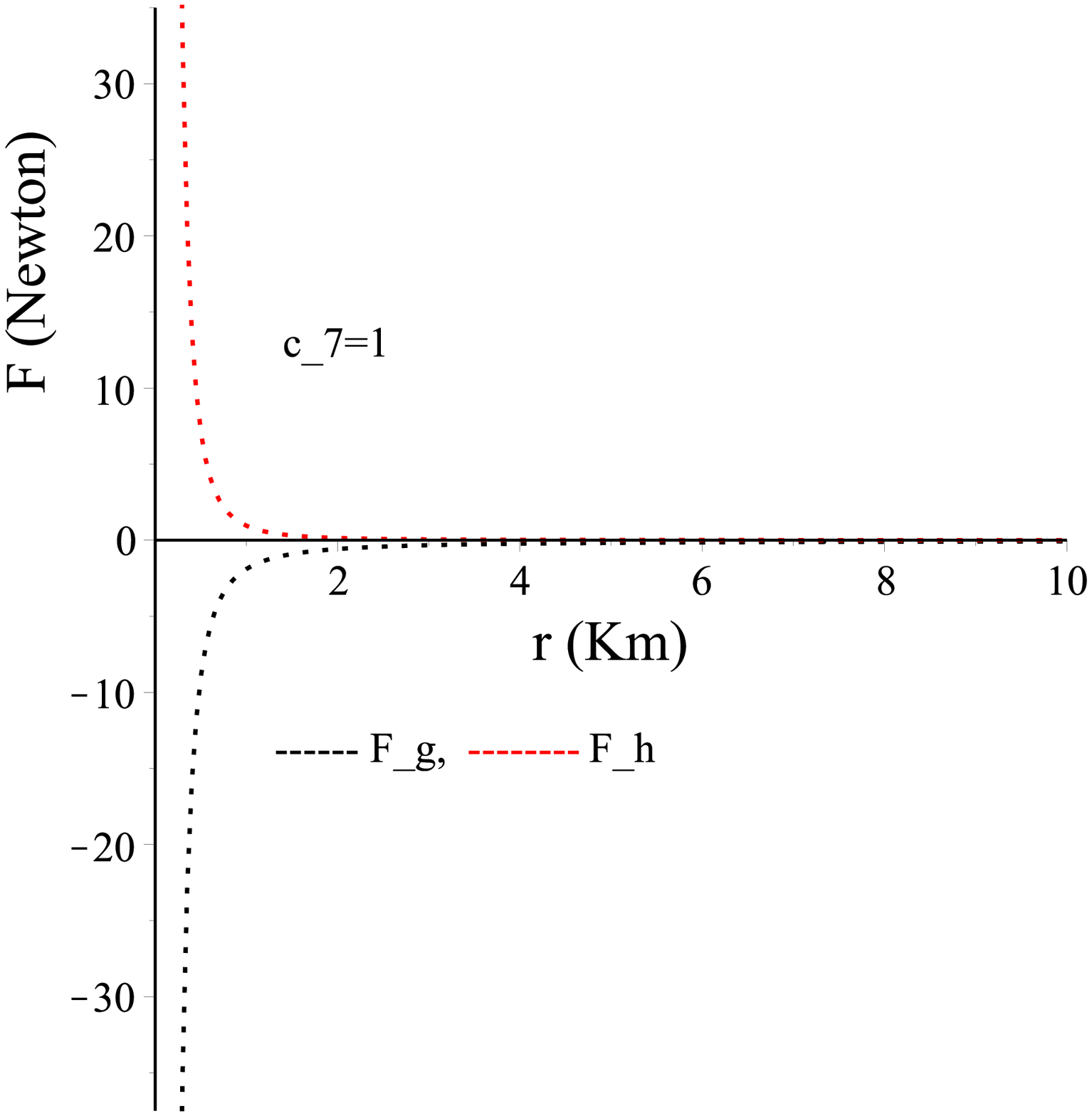}}
\caption{Two distinct forces, i.e., 
$F_g$ and    $F_h$ are
schemed via r (km) for the two models.}
\label{Fig5}
\end{figure}
\section{Conclusion and discussion}

In this study we have used two non diagonal different tetrad fields having spherical symmetry and reproduce the same associated metric. These tetrads are connected by local Lorentz transformation. We have used the CKV mechanism to reduce the highly  nonlinear partial differential equations. We have applied the field equations of $f(T)$ to the first tetrad and have obtained anisotropic system consists of four non linear differential equations. One of these deferential equations put a constraint  on the form of $f(T)$. This constraints  make the form of $f(T)$ to be $f(T)=T$. Using this form and the isotropic condition, i.e., $p_r=p_t$, we get an isotropic  solution.

For  the second tetrad we have obtained anisotropic system that consists of three non linear differential equations. We cannot solve this system without any constrains on the form of $f(T)$.  Using the constraint of $f(T)$ applied to the first tetrad, i.e. $f(T)=T$ and the condition of isotropy, we get another solution.

We have studied the physics relevant to each solution and have shown that the first  and second tetrads satisfied the energy conditions  provided that the two constants of integration involved in these solutions  be positive. We have shown that the first tetrad is not stable one  because the sound speed is negative, i.e., $\frac{dp}{d\rho}<0$ \cite{Hl}. However, the  second model has confirmed  stable manner and has shown a dynamical behavior. We have indicated that the first tetrad is not suitable to construct a stellar
model because the radius  has an imaginary quantity. In meanwhile the second model has illustrated  a stellar model that has a radius about one $Km$ and the density is not a dense on the surface. Finally we have shown that the figures of the widespread TOV equation indicate that static equilibrium has been achieved  by distinct forces. Figure 5b show that the second model has a tendency toward equilibrium  while the first one did not show such equilibrium.

\end{document}